\documentclass[%
 preprint,
superscriptaddress,
nofootinbib,
 amsmath,amssymb,
 aps,
pra,
]{revtex4-2}

\usepackage{graphicx}
\usepackage{dcolumn}
\usepackage{bm}

\usepackage{hyphenat}
\usepackage{amsmath,amsfonts}
\usepackage[hidelinks]{hyperref}
\usepackage{xcolor}
\usepackage{graphicx, tikz}
\usepackage{float}
\usepackage{amsthm}
\usepackage{tikz}
\usepackage{subcaption}
\usepackage{amssymb}
\usepackage{physics}
\usepackage[normalem]{ulem}

\theoremstyle{definition}

\numberwithin{equation}{section}

\definecolor{VPScolor}{rgb}{0.1,0.4,0.9}

\begin{document}
\title{Magic State Distillation from Entangled States}
\author{Ning Bao}
\email{ningbao75@gmail.com}
\affiliation{Computational Science Initiative, Brookhaven National Laboratory, Upton, New York, 11973}
\author{ChunJun Cao}
 \email{ccj991@gmail.com}
\affiliation{Joint Center for Quantum Information and Computer Science, University of Maryland, College Park, MD, 20742, USA}
\author{Vincent Paul Su}
\email{vipasu@berkeley.edu}
\affiliation{Center for Theoretical Physics and Department of Physics, University of California, Berkeley, CA 94720}

\begin{abstract}
    Magic can be distributed non-locally in many-body entangled states, such as the low energy states of condensed matter systems. Using the Bravyi-Kitaev magic state distillation protocol, we find that non-local magic is distillable and can improve the distillation outcome. We analyze a few explicit examples and show that spin squeezing can be used to convert non-distillable states into distillable ones. 
    Our analysis also suggests that  the conventional product input states assumed by magic distillation protocols are extremely atypical among general states with distillable magic. It further justifies the need for studying a diverse range of entangled inputs that yield magic states with high probability.
\end{abstract}

\maketitle

\section{Introduction}




Fault tolerance is an absolute necessity for leveraging the full power of quantum computing. It is known that one cannot implement a universal set of gates transversally using a single quantum error correction code~\cite{eastin09_restr_trans_encod_quant_gate_sets}. For many popular choices of codes, Clifford gates are often transversal while the non-Clifford gates are not.  Therefore, the challenge is to perform the remaining non-Clifford element, such as a T gate, fault-tolerantly. The difficulty of implementing these T gates forms a major bottleneck for the development of universal quantum computation. Various solutions for implementing non-Clifford gates have been proposed~\cite{Bravyi_2005,Knill2004,codeswitching1,Yoder17,Yoder16,stackedcode,doublecolorcode,gaugecolorcode,Brown19}. 

 One such proposal is to prepare so-called ``magic states'' and perform T-gates through measurement-based quantum computation schemes~\cite{GottesmanChuang99,Nielsen03,Leung01,Bravyi_2005} using only fault-tolerant Clifford operations. To this end, magic state distillation (MSD) protocols have been developed in order to convert product states with large overlap with the magic states into the magic states of a particular form; for the seminal work in this area, please see~\cite{Bravyi_2005,Knill2004}. However, these magic states can be costly to prepare with arbitrarily high accuracy. Such is the prevalent assumption behind the program of Clifford+T circuit decompositions and related compilation schemes~\cite{Gosset13,Amy13,Nam18,Heyfron17,Wang20}.  Because of the inefficiencies in distilling these magic states, the ability to develop a ``magic state factory'' that is capable of effectively generating such states in large quantities would be a great advance to the scalability of quantum computation.

 The conventional setup for magic state distillation takes a number of noisy ancillae, which are often represented as product of identical single qubit mixed states, and projects them onto the code subspace of a carefully chosen quantum error correction code. One then decode using a fault-tolerant Clifford circuit and obtains a number of magic states that have higher fidelity provided the inputs are above a distillation threshold. A great number of procedures and variants have been devised to various degrees of efficacy~\cite{Bravyi_2005,Knill2004,Meier12,BravyiHaah,Jones13,Jones_multilvl,Haah17}.

 While the product of identical mixed states is a natural input state, its form is rather restrictive considering the vast pool of many-body quantum states that occur naturally in physical systems that can function as magic reservoirs~\cite{Sarkar19,White:2020zoz,LiuWinter_mbmagic,Hsieh_mbmagic,haarmana}. In particular, recent progress in quantum many-body magic indicates that, magic, like entanglement, can take on various forms in such states. Indeed, while some of the magic is found locally, not unlike the conventional input states in magic distillation protocols, a large portion remains in the correlations and is distributed non-locally in entangled states. Therefore, it is natural to ask whether such ``non-local'' magic from these systems can be a) distilled and b) used to improve the outcome of distillations. More generally, it is beneficial, not only theoretically but also practically, to understand how MSD performs on a much wider class of inputs, where magic is not concentrated strictly locally on each ancilla. 

In this work, we take the first step in characterizing how more generic input states, such as entangled states, can alter the outcomes of distillation. In particular, we answer both of the above questions in the affirmative --- it is possible to distill magic that is distributed non-locally, as opposed to locally in the traditional input states. Non-local magic (NLM) can also improve the success probabilities of the distillation protocol. A wider range of input states can thus help reduce the overall cost for generating a much needed quantum resource. From a practical perspective, we show that well-known procedures used in spin squeezing that generate entanglement and non-local magic can render undistillable magic states distillable. More concretely, we conduct our explicit analysis using the Bravyi-Kitaev (BK) magic distillation protocol~\cite{Bravyi_2005} and study how it performs over different forms of entangled 5-qubit states. 

We will review some basic aspects of magic states and their distillation in Section~\ref{sec:2}. Then in section \ref{sec:3}, we construct different entangled states with varying magic distributions and perform the Bravyi-Kitaev protocol for $T$ state distillation using these as input states. 
We show that although the conventional form of input states explicitly have only single qubit magic,  states with purely non-local magic such that individual qubit reduced density matrices look non-magical can remain fully distillable. We then consider both conceptually simple and practically relevant examples where non-local magic and entanglement improve distillation outcome. This is made particularly apparent with the example of spin squeezing, where non-distillable states are rendered distillable with a slight addition of non-local magic and entanglement. 
Then in Section~\ref{sec:4}, we further analyze the properties that make certain states more distillable from the point of view of the quantum error correction code in question. 
By analyzing random distillable states, we conclude that the conventional MSD inputs are extremely atypical whereas typical distillable states have high entanglement and zero local magic. We then provide some intuitions for the patterns behind the distribution of the distillable states. Finally, in Section~\ref{sec:5}, we summarize our findings, speculations, and future directions.

\section{Background}
\label{sec:2}
\subsection{Magic}

Stabilizer states are generated by Clifford operations acting on the all zeros state. They have a wide range of applications in quantum information, condensed matter physics and quantum gravity~\cite{Gottesman_stab,Kitaev_anyon,HaPPYcode}. However, it is well known that they are not dense in the Hilbert space, and thus the Clifford operations, which can be implemented transversally for certain encoding schemes, are not sufficient for universal quantum computation. 
In fact, the Gottesman-Knill theorem~\cite{Gottesman_1998} gives us a constructive way to efficiently simulate Clifford circuits purely classically. 

However, they can be supplemented with a non-Clifford gate using measurement based protocols that only involves Clifford operations~\cite{Bravyi_2005,Knill2004} and noisy ancilla states, thus rendering the computation universal. These procedures require distilling ``non-stabilizer-ness'', or magic, into a useable form.
Therefore, magic, being a resource~\cite{veitch14_resour_theor_stabil_quant_comput} in such fault-tolerant computation schemes, aims to quantify the degree to which states fail to be captured by stabilizer states. There are many magic measures. For example, the most straightforward definition is the minimum distance between a state of interest $\rho$ and the set of stabilizer states
\begin{equation}
D(\rho) = \min_{\sigma \in \rm{STAB}}||\rho -\sigma||,
\end{equation}
 where $\rm{STAB}$ denotes the set of stabilizer states. However, this measure is difficult to compute practically. A number of computable measures have also been proposed~\cite{veitch14_resour_theor_stabil_quant_comput,howard17_applic_resour_theor_magic_states,Wang18,Beverland19}.

In this work, we will use one such measure designed for qubit systems called the robustness of magic (ROM)~\cite{howard17_applic_resour_theor_magic_states}, defined for an arbitrary density matrix \(\rho\)
\begin{equation}
    \mathcal{R}(\rho) = \min \left\{ |x|_1 : \rho = \sum_i x_i s_i \, , \, x_{i} \in \mathbb{R}\right\}
\end{equation}
where the $s_i$ correspond to density operators of stabilizer states. Note that the coefficients can be negative, allowing us to express arbitrary states in this way. This quantity is lower bounded by 1 in order to have a normalized density matrix. If \(\rho\) is a stabilizer state, then there will simply be a single non-zero coefficient.

As ROM is unity for all stabilizer states, here we use $\log(\mathcal{R})$ which we call LROM to avoid the offset for stabilizer states and for similarities with definitions like mana~\cite{veitch14_resour_theor_stabil_quant_comput}. However, unlike mana which is additive for product of magic states, LROM is, in general, subadditive

\begin{equation}
    \mathcal{R}(\rho_1\otimes \rho_2) \leq \mathcal{R}(\rho_1)\mathcal{R}(\rho_2)\implies \log(\mathcal{R}(\rho_1\otimes \rho_2))\leq \log\mathcal{R}(\rho_1)+\log\mathcal{R}(\rho_2).
\end{equation}


To distinguish local from non-local magic, we will compute magic for different subsystems in a single state. If, for instance, 
\begin{equation}
\log\mathcal{R}(\rho_{12})>\log \mathcal{R}(\rho_1)+\log \mathcal{R}(\rho_2)
\end{equation}
we will then conclude that there is magic stored non-locally in the joint system $\rho_{12}$ which is not found in each of the individual subsystems. 

\subsection{Bravyi-Kitaev Magic Distillation }
Here we review the Bravyi-Kitaev(BK) MSD protocol for distilling a particular magic state, the $T$ state. Like in the original work, we assume that stabilizer operations are easy. Therefore, $T$-type magic can be supplied by any state which obtained by a stabilizer operation acting on a specific $T$ state, which we call $\ket{T_0}$. Those familiar with the Bravyi-Kitaev protocol can mostly skim this section, though we will establish some notation that will be used later in future sections.

The $\ket{T_0}$ state is the following qubit state.

\begin{equation}
    \ket{T_0} = \cos(\theta_T)\ket{0} + e^{i\pi/4}\sin(\theta_T)\ket{1} \,\quad , \,\quad \theta_T = \frac 1 2\cos^{-1}\left(\frac{1}{\sqrt{3}}\right).
\end{equation}

It can also be advantageous to write $\ket{T_0}$ in terms of its density matrix.
\begin{equation}
\ket{T_0}\bra{T_0} = \frac{1}{2}\left(I + \frac{1}{\sqrt{3}}(X + Y + Z)\right) \, .
\end{equation}

where $X, Y$ and $Z$ are the single qubit Pauli matrices.
The density matrix makes manifest that one can also describe the state by its polarization vector $\vec{r} = \frac{1}{\sqrt{3}}(1, 1, 1)$ on the Bloch sphere. Similarly, one can write the $\ket{T_1}$ state as 
\begin{equation}
\ket{T_1}\bra{T_1} = \frac{1}{2}\left(I - \frac{1}{\sqrt{3}}(X + Y + Z)\right) \, .
\end{equation}

From the polarization vector, one can see that $\ket{T_0}$ points in the direction normal to one of the triangular faces of the stabilizer octahedron.
By the symmetries of the stabilizer octahedron, there are eight states which serve the same magical purpose, all related by single qubit Clifford operations. Thus successful $T$ distillation occurs whenever any of the eight states are produced of which $\ket{T_0}$ and $\ket{T_1}$ are two. Likewise, a $T$-state may refer to any of these states, and we define the $T$-fidelity to be the maximum fidelity of a state with any of the $T$ states,
\begin{equation}
    F_{T}(\rho)=\max_{U\in \mathcal{C}^{(2)}} \{F(\rho, U|T_0\rangle\langle T_0| U^{\dagger})\}.
\end{equation}
Here we use $\mathcal{C}^{(2)}$ to denote the Clifford group over a single qubit.

The Bravyi-Kitaev protocol for $T$-state distillation involves a recursive subroutine based on the $[[5,1,3]]$ code~\cite{Laflamme_1996} whose stabilizer group is
\begin{equation}
\mathcal{S}=\langle  XZZXI, IXZZX, XIXZZ, ZXIXZ\rangle.
\end{equation}
It distills $T$ states because the code admits a transversal implementation of the Clifford gate $B=\exp(i\pi/4)SH$ for which $|T\rangle$ is an eigenstate. As usual, $S, H$ are the single qubit $\pi/2$-phase gate and the Hadamard gate respectively.

\begin{figure}
    \centering
    \includegraphics[width=.5\textwidth]{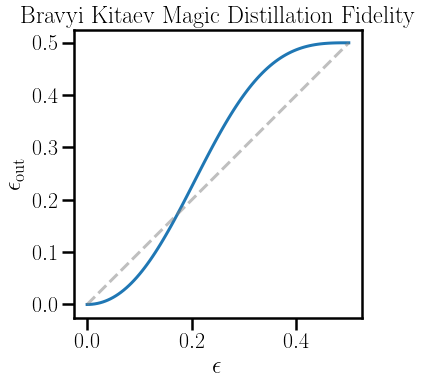}
    \caption{The output fidelity as a function of input fidelity. The cross over point is $\epsilon^* \approx 0.173$. If the input states are not sufficiently close to a $T$ state, then they will limit to the maximally mixed state. }
    \label{fig:epsilons}
\end{figure}

Given five copies of an approximate $T$ state of the following form 
\begin{equation}\label{eq:bk_input}
\rho_{in} = (1-\epsilon)\ket{T_{0}}\bra{T_{0}}  + \epsilon \ket{T_{1}}\bra{T_{1}}  \, ,
\end{equation}
the MSD protocol involves measuring the generators of $\mathcal{S}$ and post-selecting the trivial syndrome, effectively projecting the input state onto the code subspace. If a non-trivial syndrome is measured, the state is simply discarded and the protocol is restarted. The decoding circuit is then applied to return a single qubit state, which iteratively approaches a pure $T$ state, provided the fidelity of $\rho_{in}$ with any $T$-state is sufficiently high\footnote{Equivalently, it is the same to first apply the decoding circuit then measure the disentangled syndrome qubits separately~\cite{Gottesman:1997zz}.}. For every iteration of this procedure, one obtains noisy ancilla with a different fidelity. We plot the input and output fidelities as a result of this procedure in Figure~\ref{fig:epsilons}. 

It is worth noting that the input state, which is a tensor product of one qubit states, is certainly not in the code subspace. Because there is only a single logical qubit, one can decompose the projected state in the basis of the encoded magic states $\ket{T^{L}_{0}}$ and $\ket{T^{L}_{1}}$. 
The crux of the analysis lies in showing that for sufficiently small $\epsilon$, the $T$-fidelity increases. If for any $\epsilon<\epsilon^*$ such that the protocol always increases the $T$-fidelity of the input state, then $1-\epsilon^*$ is known as the distillation threshold.

When it comes to the efficiency of the procedure in creating magic states, there are two factors that determine the resource estimate. Assuming there is a target $T$-fidelity, one can use the curve in Figure~\ref{fig:epsilons} to estimate how many rounds of distillation are required starting from an initial pool of noisy ancilla. The second quantity of practical importance is the success probability for measuring trivial syndrome. By success probability we mean the probability of measuring the trivial syndrome corresponding to a single iteration of the distillation, as opposed to producing an approximate $T$ state with the desired final fidelity after multiple rounds.

For input states of the form \eqref{eq:bk_input}, the success probability depends on $\epsilon$ as well. As seen in Figure~\ref{fig:bk_round}, this is upper bounded by 1/6, meaning every round of distillation in expectation requires at least 30 approximately magical qubits. Since it is assumed that all 5 input qubits are the same, one actually needs access to an exponential number of noisy ancilla due to the recursive nature of the protocol.


The utility of $T$ states was also discussed in the same paper by Bravyi and Kitaev.
They showed explicitly how to enact a non-Clifford gate with access to $T$ states and only Clifford operations. The fact that the same state was the fixed point of the distillation procedure and enables Clifford operations to enact universal quantum computation was the reason for calling these states ``magic states''.

\section{Magic Distillation Beyond Product States}
\label{sec:3}
In this section, we investigate whether existing MSD procedure can distill magic from non-product states. Indeed, the conventional input for such protocols $\rho_{\rm in}^{\otimes n}$ is a product state. Magic is localized to each bit and there is clearly no magic hidden in the quantum correlations.
However, one might ask whether other forms of multi-qubit magic states can also be used to distill desired magic states as long as certain conditions are satisfied. In particular, can the distillation procedure distill magic from states where the state has entanglement or correlation such that magic is non-locally distributed, i.e. magic of any single qubit is small whereas that of a larger subsystem is non-trivial? For example, such states can exist in the ground state of  quantum many-body systems~\cite{White:2020zoz, Sarkar19}. To this aim, we construct a few informative examples below each with distinct but carefully designed magic distributions and answer the above question in the affirmative by testing them using the BK protocol. 
For all of the distillation procedures below, we only use the entangled states as inputs for the first iteration. The ensuing iterations works under the conventional assumptions where multiple tensor copies of the outputs from earlier iterations are used.

\subsection{Phase-GHZ and GHZ-T states}\label{ssec:3A}

\begin{figure}
    \centering
    \includegraphics[width=0.44\textwidth]{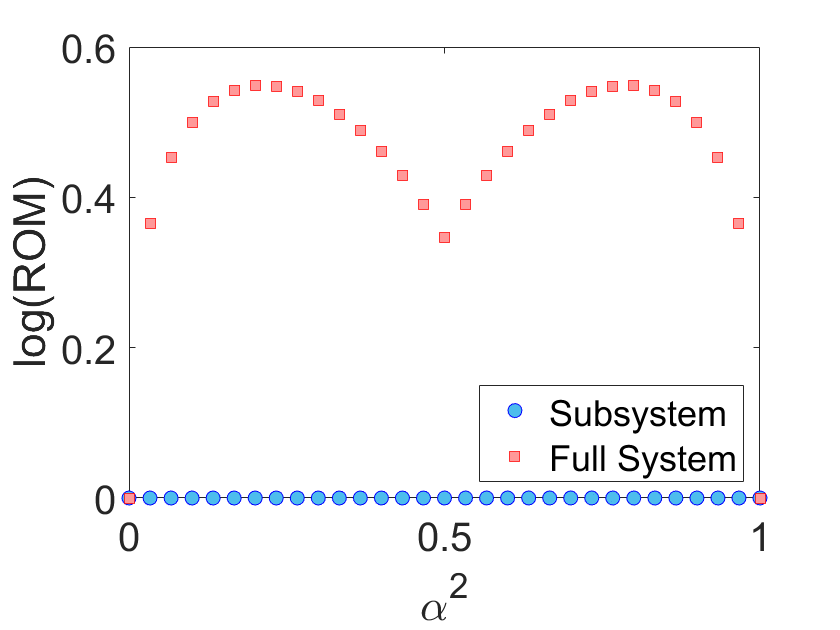}
    \includegraphics[width=0.54\textwidth]{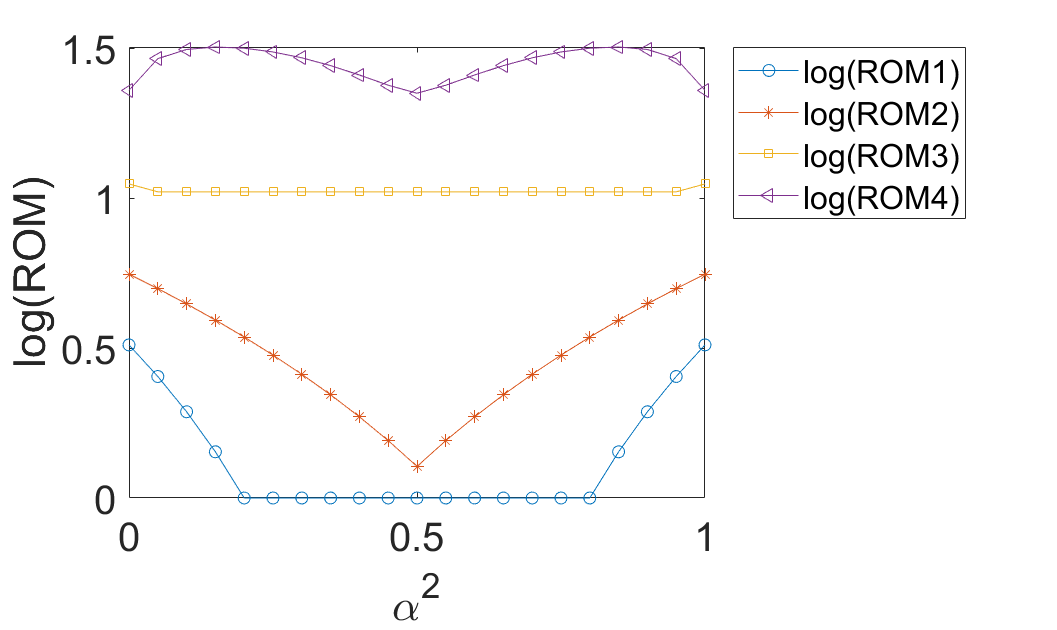}
    \caption{(a) Magic distribution of the Phase-GHZ state with phase $\phi=\pi/4$. For PG, all subsystems have the same form of reduced density matrix. (b) Magic of various subsystems for the 4-qubit GHZ-T state as a function of $\alpha^2$. }
    \label{fig:pg_gt_magic}
\end{figure}

We begin with a few structurally simple states to build intuition. Consider GHZ states in the computational basis with a phase. 

\begin{equation}
    \ket{PG(\alpha)}_{n} = \alpha \ket{0}^{\otimes n} + e^{i\phi}\beta \ket{1}^{\otimes n} .
\end{equation}
For instance, one can choose $\alpha\in [0,1]$ which sets the relative weighting of the superposition. By normalization, we choose $\beta = \sqrt{1 - \alpha^2}\in \mathbb{R}$. 
 For extreme values of $\alpha/\beta$, this becomes arbitrarily close to a simple product state, allowing us to ``tune'' the presence of magic. The specific magic distribution is shown in Figure~\ref{fig:pg_gt_magic}a. 

This state provides a steep contrast with the conventional input states of MSD --- no proper subsystem is magical as they are all density matrices diagonal in the computational basis. However, the overall state clearly is magical because one can turn this state into the product of $|0\rangle$s and a single qubit magic state $\alpha|0\rangle+e^{i\phi}\beta|1\rangle$ by applying a series of Clifford gates.

\begin{figure}
    \centering
\includegraphics[width=.85\textwidth]{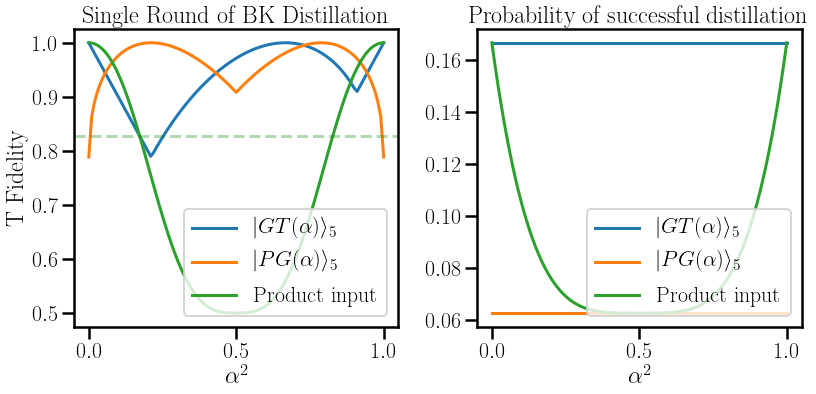}
\caption{With our choice of simple entangled states, we apply the BK protocol and measure the magic fidelity of the resulting qubit. For comparison, we also compute it for the product states considered in the original work. Notably, the output fidelity is above threshold across a wide range of $\alpha^2 \sim 1-\epsilon$. The output of this round can then be recursively iterated. The green line shows the distillable threshold for states of the form \eqref{eq:bk_input}. For the other states, they reach a pure $T$ state except a few specific values of $\alpha$. This more lenient threshold is not a contradiction since the output density matrix of this round is not of the same form. On the right, we show the probability to measure the trivial syndrome. To actually produce a $T$ state, one needs the procedure to both converge to a $T$ state and measure the trivial syndrome. }
    \label{fig:bk_round}
\end{figure}

This state is distillable using BK (Figure~\ref{fig:bk_round}) for most values of $\alpha$ because the output is above the distillation threshold (dashed line). Although BK is suboptimal for distilling magic from such kind of states, it clearly demonstrates that non-local magic remains distillable using existing protocols.

Other than the conventional product input states which have magic strictly localized to each qubit and the Phase-GHZ state above where magic is strictly global, one can also construct states that have magic at different scales. Consider a state that takes the following form on $n$ qubits \footnote{While we do not use the degree of freedom of the relative phase in this work, it is one that's available to us and others for future work.}.
\begin{equation}
    \ket{GT(\alpha)}_{n} = \alpha \ket{T_0}^{\otimes n} + e^{i\phi}\beta \ket{T_1}^{\otimes n} .
\end{equation}
Such a state is chosen precisely because it is a combination of the noisy ancilla input and the Phase-GHZ state above. While each one-qubit reduced density matrix mimics the conventional noisy ancilla input that shows up in the protocol (e.g. $\epsilon = \beta^2$), the overall state is also similar to the Phase-GHZ state where there is additional magic globally. Thus for $\alpha$ or $\beta$ small, magic is locally distributed. However, for other choices of $\alpha,\beta$ the state can pick up a non-trivial contribution from non-local magic in the $n$-qubit GHZ state. One can see that magic (Figure~\ref{fig:pg_gt_magic}b) on small subsystems of one or two qubits has a distribution similar to that of the conventional input states. However, as we go to larger subsystems, there is also more non-local magic which adds to the total global magic available in the state.

Magic in such states are mostly distillable. See Figure~\ref{fig:bk_round}. However, the dips in the distillability plot now no longer coincide with the locations where there is less total magic, which was the case for Phase-GHZ states. 

\begin{figure}
    \centering
    \includegraphics[width=.8\textwidth]{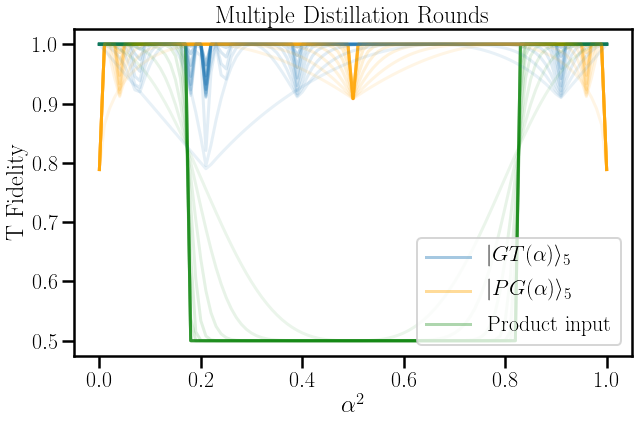}
    \caption{Multiple iterations of the BK protocol for simple input states. Over a wide range, our choice of toy entangled states are distillable. Original product input states are limited to the tail ends where $\alpha^2 < \epsilon_*$ or $\alpha^2 > 1 - \epsilon_*$.}
    \label{fig:bk_rounds}
\end{figure}

We can ask whether how global magic helps with distillation in these examples. Other than the clear improvement in distillability in the region where $\beta^2=\epsilon \in[\epsilon^*, 1- \epsilon^*]$, where the conventional inputs are completely undistillable, there is also other improvements in the form of success probability, i.e., the probability of measuring the trivial syndrome~(Figure~\ref{fig:bk_round}). 
Indeed, the GHZ-T state, which has single qubit states identical to conventional noisy ancilla input but have non-zero non-local magic retains a relatively high success probability for all values of $\alpha$, which is distinct from the product state counterpart.

\subsection{Squeezed Spin States from Noisy Ancillae}
\label{subsec:sss}
The above examples are conceptually clean and easy to understand, but may be hard to prepare in the laboratory. For entangled input states that are slightly more realistic, we consider initially states that are product stabilizer states that  suffer from single qubit coherent noise. We then perform a global entangling unitary usually used in spin squeezing procedures~\cite{SSS93} on such states and examine how it improves distillation outcome.

The initial ancilla states before squeezing are spins that are parallel to, say, the x-axis, but have small misalignments across the spins. More explicitly,
\begin{equation}
    |\psi_{\rm ini}\rangle = \bigotimes_{i=1}^5 |\theta_i\rangle,~~~~|\theta_i\rangle = \cos\theta_i|0\rangle+\sin\theta_i|1\rangle,
    \label{eqn:mis_spin_state}
\end{equation}
where the orientation of each spin $\theta_i$ can fluctuate to simulate inhomogeneity of the initial spin alignment. This is modelled by choosing each $\theta_i\in [\bar{\theta_i}-\theta_{\rm max}, \bar{\theta_i}+\theta_{\rm max}]$ randomly from a uniform distribution. We take the means $\bar{\theta_i}$ to be equal for all $i$. For example, $\bar{\theta_i} =\pi/4$ creates a product state that is approximately $|+++++\rangle$ but with small misalignment across the spins. Despite the small amount of local magic introduced to $|\psi_{\rm ini}\rangle$ through the spin misalignments, such states remain mostly undistillable, especially for $\theta_{\rm max}$ small.



\begin{figure}
    \centering
    \includegraphics[width=0.49\textwidth]{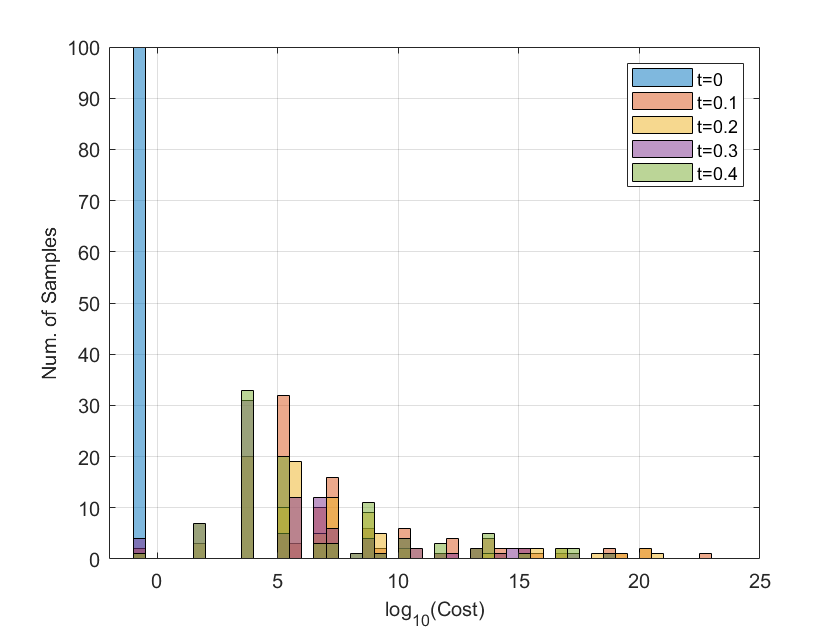}
    \includegraphics[width=0.49\textwidth]{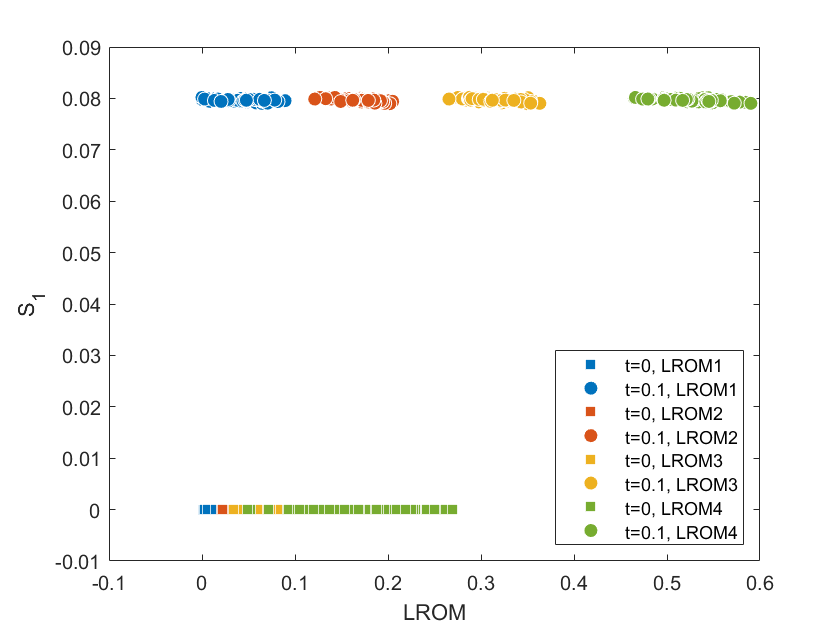}
    \caption{(a)Figure of distillability of states with ($t\ne 0$) vs without ($t=0$) squeezing ($\theta_{\rm max}=0.05$). States without squeezing (blue) are undistillable and are assigned cost $-1$. (b) Single site entanglement and magic of various subsystems of squeezed (colored disks) vs unsqueezed state (colored squares) with $ \bar{\theta}_i=\pi/4$.}
    \label{fig:SSS_distill}
\end{figure}



To define the unitary squeezing process, consider a total spin-$5/2$ system consisting of 5 qubits. It is known that a pure product state of qubits is not a squeezed state~\cite{SSS93,SSS_entanglement} and squeezing introduces entanglement among the qubits.
For example,
\begin{equation}
    |+++++\rangle = 2^{5/2}\sum_{k=0}^{5}\sqrt{{5 \choose k}}| S=\frac 5 2, S_z=\frac 5 2-k\rangle
\end{equation}
Squeezing can then be performed by considering some unitary rotation on the $S=5/2$ submanifold. The initial spin misalignments  sometimes introduce components that are outside the $S=5/2$ submanifold, and the unitary mapping over the 5 qubits leave these components unchanged. We can use the one-axis twisting method~\cite{SSS93} with unitary $U(t)$ to generate a number of squeezed states $|\psi_{\rm fi}\rangle$
\begin{equation}
    U(t)=\exp[-it S_z^2]\oplus I, ~~~~ ~|\psi_{\rm fi}(t)\rangle = U(t)|\psi_{\rm ini}\rangle.
    \label{eqn:oneaxis}
\end{equation}
Note that although $U(t)$ is not a Clifford unitary, there is also no need for it to be fault-tolerant or extremely precise as a range of different $t$'s can all yield similar outcomes.

For small $t$, the action of $U(t)$ introduces a small amount of mostly non-local magic and entanglement to the initial state.  See Figure~\ref{fig:SSS_distill}b for results when $\langle \theta_i\rangle =\pi/4$. 
 Surprisingly, the previously undistillable initial states are now mostly distillable by applying $U(t)$ (Figure~\ref{fig:SSS_distill}a). Here cost is defined as the average number of copies of the same state needed to distill a single copy of the $T$ state with a reasonably high fidelity ($F_T(\rho)\geq 0.97$ for the plots). If the resulting output does not exceed the prescribed fidelity cut off after 15 iterations, we deem the state undistillable and it is assigned cost $-1$ to differentiate from the rest of the samples.
 

 This is another clear demonstration that entangled states generated in a more practical laboratory setting~\cite{SSS_experiment} can help enhance the outcome of magic state distillation.
While this is encouraging, squeezing can also act like a source of correlated noise and does not always improve distillation outcomes. See Appendix~\ref{app:SSS} for further discussions on squeezed spin states and magic state distillation. 


\section{Understanding Distillability}
\label{sec:4}

In this section, we examine the underlying $[[5,1,3]]$ code that powers the BK distillation procedure to understand properties of states that contain distillable magic. In doing so, we present a few key observations. First, using elementary ideas from quantum error correction, we can deduce the form of non-local magic content for the ideal input state. We then investigate these intuitions by generating random BK-distillable states whose success probability is lower bounded by that of product input states. Studying their entanglement and magic distribution properties tells us that the typical product ansatz is highly fine-tuned. Second, we can explain explicitly why the total magic in a state does not correlate with the distillable magic for this protocol. We expect the lessons here to generalize appropriately for any magic distillation procedure which relies on a quantum error correcting code.

As mentioned previously, the Bravyi-Kitaev distillation procedure (decoding followed by postselecting the trivial syndrome) can be represented as a linear map $M:\mathcal{H}_{\rm in}\rightarrow \mathcal{H}_{\rm out}$, where $\mathcal{H}_{\rm in}, \mathcal{H}_{\rm out}$ are 32 and 2 dimensional Hilbert spaces, respectively. Then for each normalized 5-qubit input $|\psi_{\rm in}\rangle \in \mathcal{H}_{\rm in}$, we can obtain a single-qubit output $|\psi_{\rm out}\rangle\in \mathcal{H}_{\rm out}$ such that

\begin{equation}
    \lambda|\psi_{\rm out}\rangle = M|\psi_{\rm in}\rangle.
\end{equation}
As a portion of $|\psi_{\rm in}\rangle$ may not lie in the code subspace, the probability of success $|\lambda|^2$, i.e. measuring a trivial syndrome, satisfies $0\leq |\lambda|\leq 1$. It is clear that states with more overlap with the code subspace have higher
success probability.

Working in the basis of the code subspace, any input state with unit
success probability can be written as 
\begin{equation}
    |\psi_{\rm in}\rangle = a|{T}^L_0\rangle +b|{T}^L_1\rangle
\end{equation}
where the size of $a$ will determine the output state fidelity. This works because the perfect code encodes a single qubit.

It is immediately clear that states close to the encoded $T$ states will have both high success probability and good $T$-fidelity. Furthermore, because the code corrects two qubit erasure errors, any encoded information must be found in 3 or more qubit subsystems. Therefore, for states with high success rate, i.e., large overlap with the code subspace, the amount of magic found in 3 or more qubit subsystems should positively correlate with the output fidelity. On the other hand, such states will have very little magic in one or two qubit subsystems.

However, when a state does not overlap significantly with the code subspace, namely $|\lambda|$ small, the connection between the amount of magic, its distribution and output fidelity is no longer clear.

To get an idea of a typical distillable state, we generate random states uniformly in the space of states where their output fidelity is above the BK distillation threshold and   $|\lambda|>0.16$ is reasonably large. As a reference, the
success probability for product input states close to $T$ states is $\sim 0.167$ (see Figure~\ref{fig:bk_round}).

\begin{figure}%
    \centering
    \subfloat[\centering ]{{\includegraphics[width=0.45\textwidth]{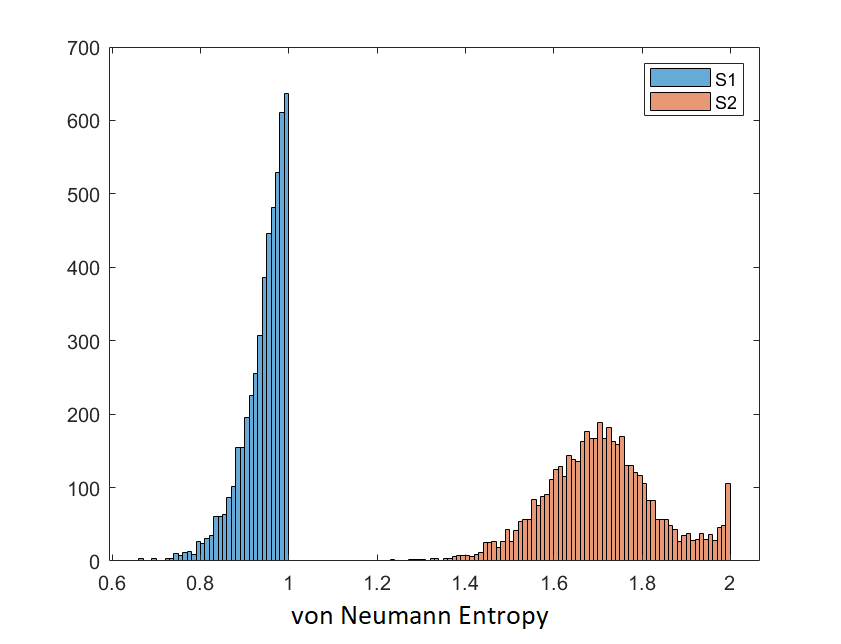} }}%
    \qquad
    \subfloat[\centering ]{{\includegraphics[width=0.45\textwidth]{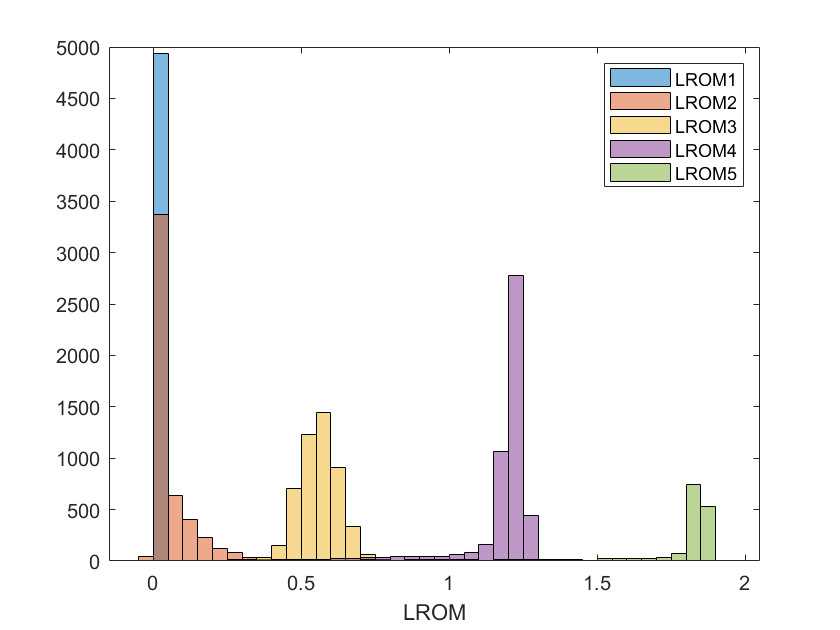} }}%
    \caption{(a) Distribution of von Neumann entropies for single and two qubit subsystems. (b) LROM distribution in the random BK-distillable states.}%
    \label{fig:RN_vNROM_distr}%
\end{figure}

We see that the typical distillable states are not only entangled, but also have non-local magic (Figure~\ref{fig:RN_vNROM_distr}).
As we see, any subsystem is close to being maximally mixed while magic localized to any single qubit is virtually non-existent. 
This is in sharp contrast with the conventional input states which have neither. Therefore, this observation indicates that having many-body quantum states beyond the typical product input states can improve the magic distillation outcomes.

However, unlike the conventional input states, there is no obvious correlation between the total amount of magic or entanglement a state has and the distillation outcome. In fact, if anything, too much magic can negatively impact distillability in terms of the
success probability. We can try to understand this observation from two limits. 

\begin{figure}
    \centering
    \includegraphics[width=0.7\textwidth]{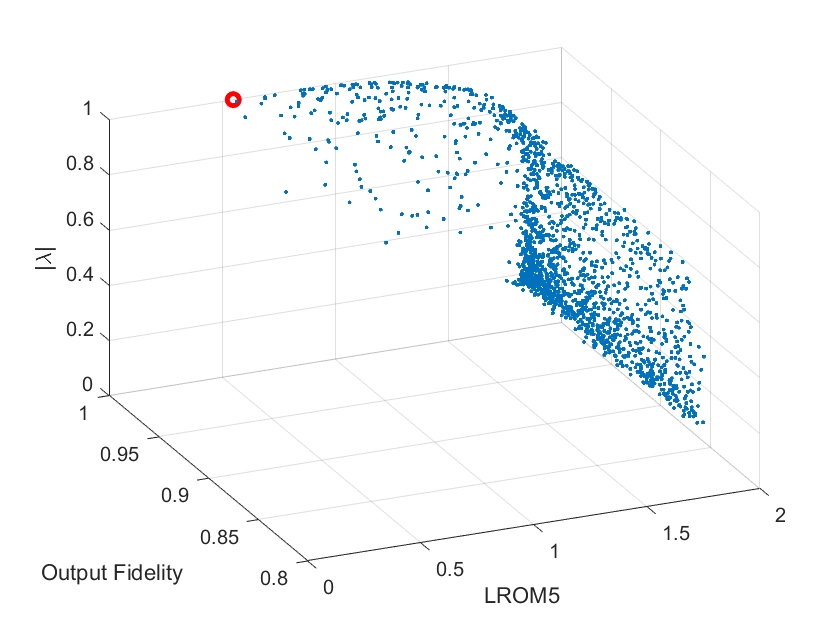}
    \caption{LROM5 of random states (blue) against output fidelity and
    success probability. Red circle denotes the location of the encoded $T$ state. }
    \label{fig:lrom5_RN}
\end{figure}

\begin{figure}
    \centering
    \includegraphics[width=0.7\textwidth]{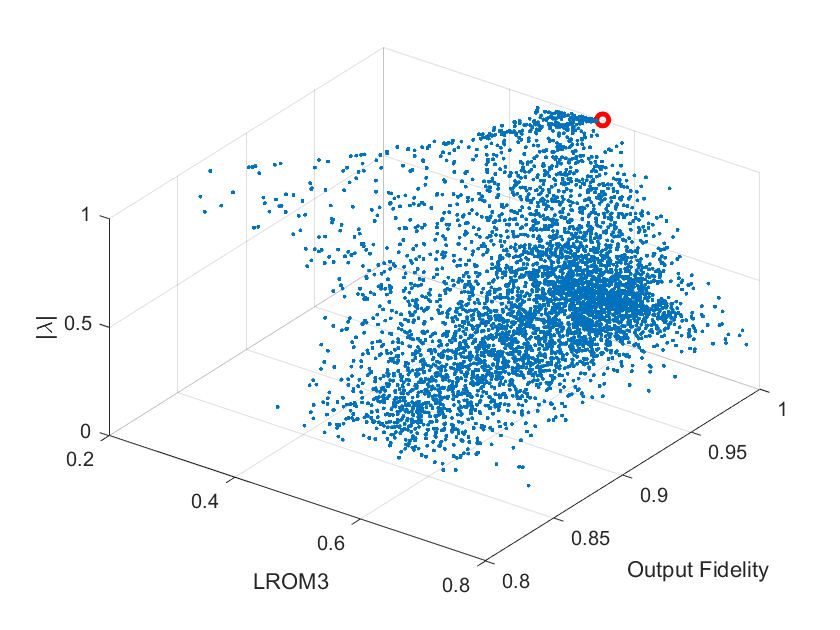}
    \caption{LROM3 of random states (blue) against output fidelity and
    success probability. Red circle denotes the location of the encoded $T$ state. On the plane $|\lambda|=1$, there is clear correlation between LROM and output fidelity. }
    \label{fig:lrom3_RN}
\end{figure}

In the limit where the state is perturbatively close the an encoded $T$ state, then having a state whose magic or entanglement pattern that deviates from the state that has the single best distillable outcome will obviously  result in a worse distillation outcome. The maximum amount of magic for a state in the code subspace is upper bounded by that of a single $T$ state ($\mathrm{LROM} \sim 0.55$). If the input state has more magic than this, then it is necessarily out of the code subspace thus resulting in a lower
success probability~(Figure~\ref{fig:lrom5_RN}). If, on the other hand, the total magic is less than the upper bound, then it can have a large overlap with the code subspace. In that case, the
success probability is close to 1, but its output fidelity must suffer for being farther from the encoded $T$ state. There is a direct correlation between the amount of magic and output fidelity. This can be seen in Figure~\ref{fig:lrom3_RN}, but is less pronounced in Figure~\ref{fig:lrom5_RN}.
Interestingly, it is also difficult to find a distillable state that has less total magic than that of a $T$ state but also small overlap with the code subspace. This is likely related to the fact that (Haar) random states are magical~\cite{haarmana}\footnote{Interestingly, LROM of a larger subsystem has such a smaller variance compared to that of a smaller system. This is also qualitatively consistent with results from Haar random states~\cite{haarmana} where the variance of magic scales roughly as $\sim e^{-L}$ where $L$ is the number of sites.}.  

In the limit where a state has small overlap with the code subspace ($|\lambda|\lesssim 0.4$), the previous arguments no longer apply. Instead, there is no clear correlation between output fidelity, success probability, and the amount of magic. 

One can also compute the average cost of distilling magic from these typical states. Again, there is no obvious correlation between the total magic of a state and the total cost for generating a magic state with desired $T$-fidelity $F_T(\rho)>0.999$.

\begin{figure}
    \centering
    \includegraphics[width=0.8\textwidth]{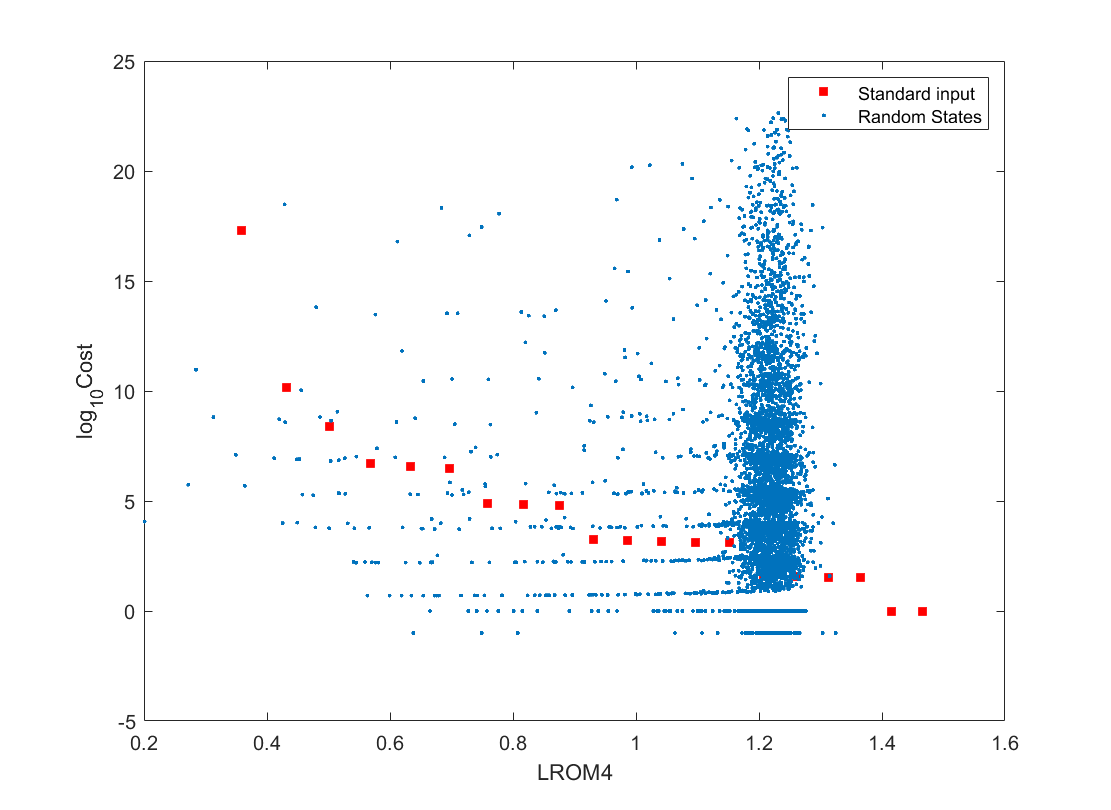}
    \caption{Plotting LROM4 and cost for random states (blue) and conventional inputs (red). Cost is set to $-1$ if the state does not exceed the prescribed fidelity $F_T(\rho)>0.999$ within 15 iterations of the protocol.}
    \label{fig:cost}
\end{figure}

However, there are states with magic lower than that of a typical state that have slightly lower cost, as shown in the thin narrow bands stretching to the left of the cluster (Figure~\ref{fig:cost}). These are precisely the states that have higher
success probabilities which are closer to the encoded $T$ states. Curiously, these states concentrate on the ridges but are not found anywhere in-between, unlike the typical states that contain more magic.



\section{Discussion}
\label{sec:5}


In this work, we have examined how entanglement and non-local magic can impact the outcome of magic state distillation. Using the BK protocol on 5 qubits, we establish that non-local magic is indeed distillable. Furthermore, numerical experiments indicate that non-local magic in these entangled states can also ``assist'' the usual magic distillation process by lowering the cost of overall distillation. This is outlined conceptually by studying the GHZ T and GHZ phase states. More practically, we have also constructed a spin-squeezing-assisted procedure for noisy ancilla states. It is shown that certain squeezing operations indeed render non-distillable states distillable, which may be more relevant in an experimental setting.

Although the overall impact of many-body magic is still largely unclear in the context of distillation, we build up some generalizable intuition using the 5-qubit code and argue that for the existing distillation protocols based on quantum error correction codes, a typical distillable state should be entangled. 

Because of the common assumptions behind magic state distillation, one might have suspected that entanglement generally introduces a source of correlated noise, thereby reducing the output fidelity and success rate of magic distillations. However, this expectation is likely a streetlight effect for only searching perturbatively near the conventional product input states. While it is a valid expectation in certain cases (e.g. Figure~\ref{fig:nearT_distill}), it is abundantly clear that by analyzing the entanglement structure and magic distribution of the encoded $T$ state, a typical distillable state is anything but a product state with local magic. 

One might also suspect that a larger amount of total magic would generally lead to better distillation outcomes. This is somewhat suggestive from Figures~\ref{fig:pg_gt_magic}a and \ref{fig:bk_rounds} for Phase GHZ states. However, this correlation is only manifest when the input lies predominantly in the code subspace. Analysis in Section~\ref{sec:4} indicates that this intuition has limited applicability when the overlap between the input and the code subspace is small. There it is unclear what properties would positively correlate with distillability.

Since our work suggests that entangled states may assist magic distillations, it opens up a number of directions in the area of many-body magic~\cite{Sarkar19,White:2020zoz,LiuWinter_mbmagic,haarmana}. Thus far we have only focused on examples and numerical analysis, however a more systematic approach is needed to understand many-body magic distillability that is not protocol specific. Generally, we want to understand the features in the distillation protocol and those of the many-body state that will lead to better distillation outcomes. In particular, it opens up the question of what a ``typical'' distillable magic state should look like if one looks beyond product of identical states. It is also unclear whether existing protocols are optimal for distilling many-body magic, as the input states are drastically different. For example, while it is well-known that bound magic states exist~\cite{CampbellBrowne_boundmagic} assuming identical inputs, it is not clear to what extent the same conclusion applies when one also considers more general input states. We would also like to understand what kind of protocols are near-optimal in distilling magic from condensed matter systems such as the low energy states of a spin-chain where magic is abundant. 

It is possible that holographic quantum error correction codes may also be useful in this regard for distilling multi-scale non-local magic from CFTs due to its multi-scale, self-similar structure. It is pointed out recently~\cite{Cree2021} that holographic codes with good complementary recovery properties have difficulties implementing transversal non-Clifford gates. Nevertheless, it is still entirely possible to construct holographic codes in general that do have transversal $T$ gates. For instance, this can be done by using codes with transversal non-Clifford gates\cite{csstranst,zeng07} such as the $[[15,1,3]]$ CSS code~\cite{1996Knill,Anderson2014} as the base tensor, which supports a transversal logical $\bar{T}$ gate. Using the general methodology outlined in~\cite{cao2021} (and more comprehensively in~\cite{CaoLackey:2021}) for building tensor networks on 2d hyperbolic space with regular tessellations, one can produce a holographic code constructed using the $\{15,q\geq 4\}$ tiling. This results in an $[[n,k]]$ CSS code that admits a transversal $\bar{T}^{\otimes k}$ gate\footnote{Note that the tensors or codes used to build the holographic code can be the same up to local Clifford equivalences.}. The same methodology also applies to more general $[[2^{n-1},1,3]]$ punctured Reed-Muller codes~\cite{puncturedReedmuller} in a $\{2^{n-1},q\geq 4\}$ tessellation. The resulting stabilizer code would again admit a transversal non-Clifford gate. It should also be noted that transversal non-Clifford gates are not always necessary for distillation. Indeed, the 5 qubit code does not admit a transversal T gate, however, the transversal $B$ gate enables distillation of $T$ states. A holographic code that serves a similar purpose is the $[[n,k]]$ stabilizer code known as the HaPPY pentagon code~\cite{HaPPY}. There  the logical $\bar{B}^{\otimes k}$ is transversal and a protocol similar to BK may be sufficient for $T$ state distillation. 



\section*{Acknowledgements}
We thank Greg Bentsen for the stimulating discussions on the experimental feasibility of creating entangled states and the suggestion to look at spin squeezing. C.C. acknowledges the support by the U.S. Department of Defense and NIST through the Hartree Postdoctoral Fellowship at QuICS, by the Simons Foundation as part of the It From Qubit Collaboration, and by the DOE Office of Science, Office of High Energy Physics, through the grant DE-SC0019380. N.B. is supported by the Department of Energy under grant number DE-SC0019380, by the Computational Science Initiative at Brookhaven National Laboratory, and by the U.S. Department of Energy QuantISED  Quantum Telescope award. 

\bibliographystyle{unsrt} 
\bibliography{references} 

\appendix
\section{Distillability of various Phase-GHZ and GHZ-T states}
\begin{figure}
    \centering
    \includegraphics[width=.45\textwidth]{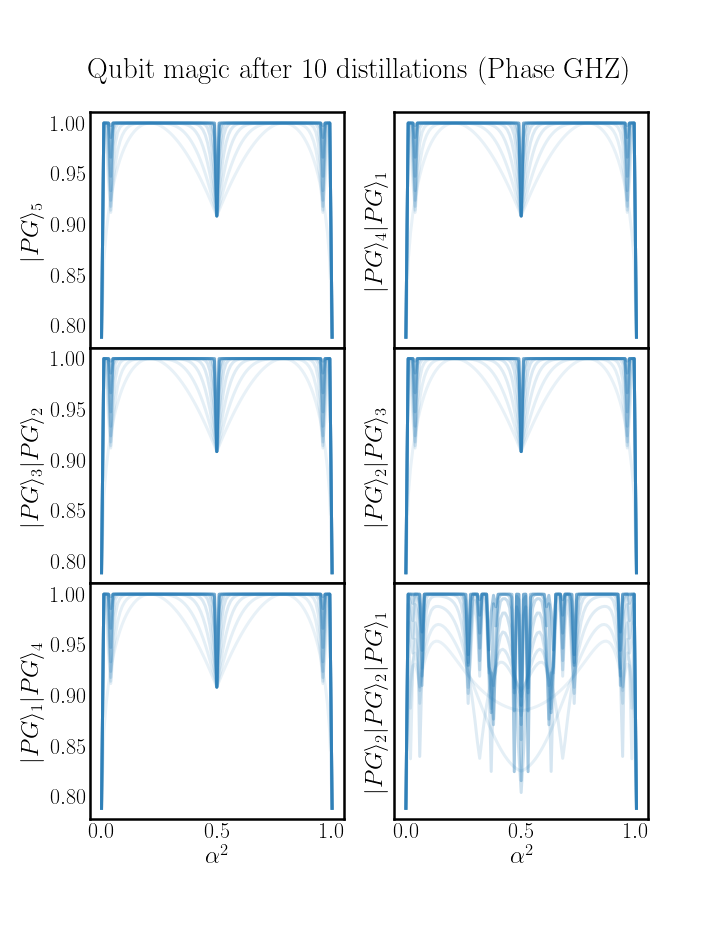}
    \includegraphics[width=.45\textwidth]{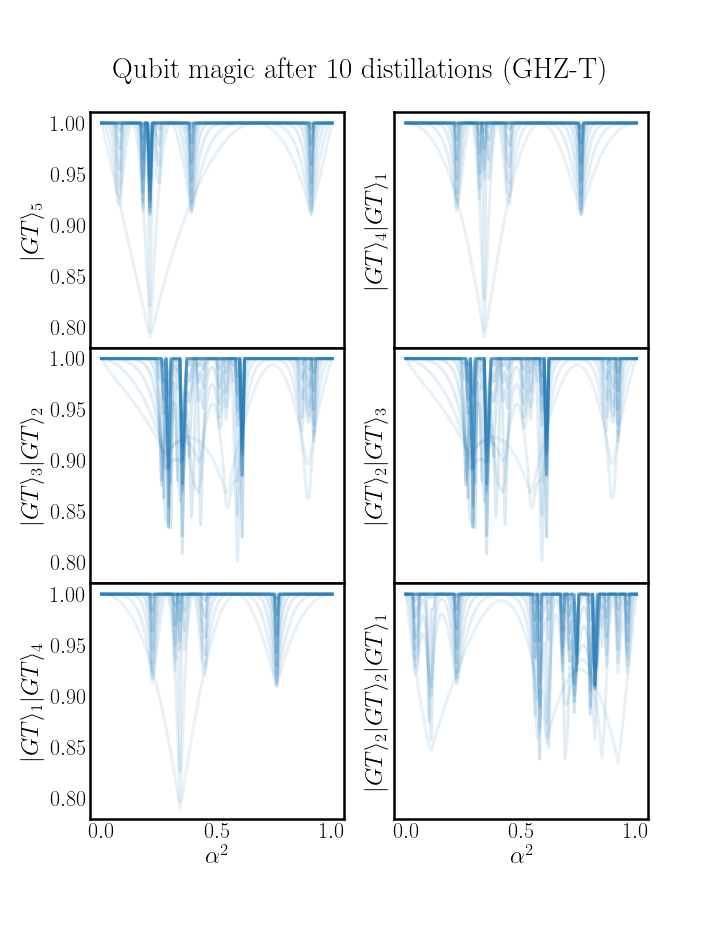}
\caption{Distillability of composite $\ket{PG}_k$ and $\ket{GT}_k$ states, where $\alpha$ is taken to uniform across individual components. We plot the magic fidelity of the qubit upon a successful distillation round with darker curves corresponding to later rounds. The darkest line is the resulting $T$-fidelity after 10 rounds. Because $\ket{PG}_k$ and $\ket{GT}_k$ contain the same amount of magic for all values of $k$, composite 5 qubit states with more individual states have more total magic in the system. We find that the presence of more total magic does not generally enhance distillability for the BK protocol.}
    \label{fig:PG_GT_combo}
\end{figure}

\begin{figure}
    \centering
    \includegraphics[width=.45\textwidth]{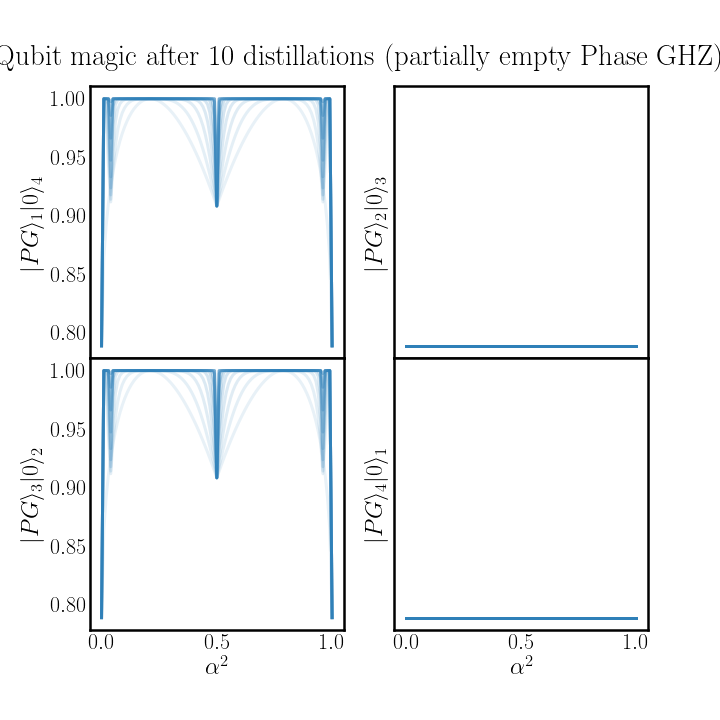}
    \includegraphics[width=.45\textwidth]{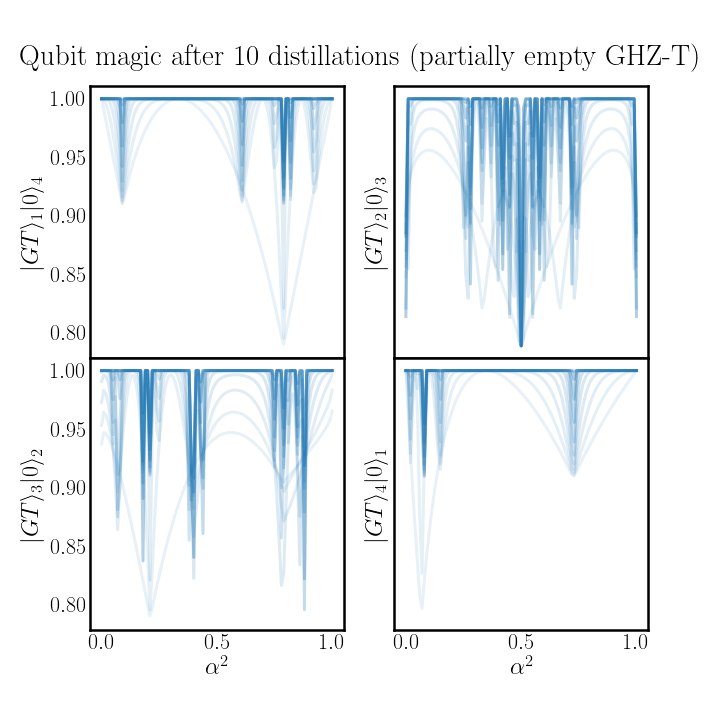}
\caption{Distillability of composite $\ket{PG}_k$ and $\ket{GT}_k$ states, with some qubits in the $\ket{0}$ state. A notable finding is that one can get similar distillability without having to entangle the full 5 qubit state. }
    \label{fig:PG_GT_select}
\end{figure}

In this appendix we examine the distillability of states similar to those presented in \ref{ssec:3A}, the Phase GHZ and GHZ T states. As a reminder, these were $n$ qubit states that took the following form.
\begin{equation}
    \ket{PG(\alpha)}_{n} = \alpha \ket{0}^{\otimes n} + e^{i\phi}\beta \ket{1}^{\otimes n} \, , \, 
    \ket{GT(\alpha)}_{n} = \alpha \ket{T_0}^{\otimes n} + e^{i\phi}\beta \ket{T_1}^{\otimes n} 
\end{equation}
Because the Bravyi-Kitaev protocol takes as input a 5 qubit state, it was a natural choice to pick $n=5$. Not only does this make each of the reduced density matrices the same, the input state is symmetric between the five qubits. Here, we now consider a more broad construction where we make a 5 qubit state out of a combination of smaller $\ket{PG}_n(\ket{GT}_n)$ states. This may lead to easier experimental implementation if it is much easier to generate, say, $\ket{PG}_2$ and $\ket{PG}_3$ than $\ket{PG}_5$. On the theoretical side, it is interesting to ask whether the presence of more magic yields better distillation outcomes. This is because $\ket{PG}_n$ has the same magic regardless of $n$, so $\ket{PG}_2\ket{PG}_3$ would have twice as much global magic as $\ket{PG}_5$ for the same $\alpha$.

We consider two primary variations. The first is where we comprise the 5 qubit state of multiple $\ket{PG}_k$ states such that $\sum k = 5$. We make a few observations. The first is that adding more total magic does not seem to enhance the ability to distill. This lines up with the expectations laid out in Section~\ref{sec:4}. Interestingly, all of these states remained distillable for generic $\alpha$. 

The second variation allows for some of the qubits to be in the $\ket{0}$ state in the computational basis, a consideration for experimental setups where the $\ket{0}$ state is effectively just resetting a clean qubit. Clearly, these qubits do not add to the available magic. The results are displayed in Figure~\ref{fig:PG_GT_select}. With the exception of Phase-GHZ states with even numbers, these states also have a wide range of distillability.
 
\section{More on Spin Squeezing and Distillation}
\label{app:SSS}
Although a global entangling unitary like the one defined in Section~\ref{subsec:sss} can improve the distillation outcome, this improvement will depend on a number of factors and is not universal. Here we analyze how different initial states modified the same squeezing unitary $U(t)$ may impact the distillation outcome. Then we examine the effect of a different squeezing unitary $V(t)$. We find that improved distillation is unlikely to be connected to squeezing and overall amount of entanglement. However, its success may be related to the non-local magic injected during the squeezing process.

\begin{figure}[H]
    \centering
    \includegraphics[width=0.7\textwidth]{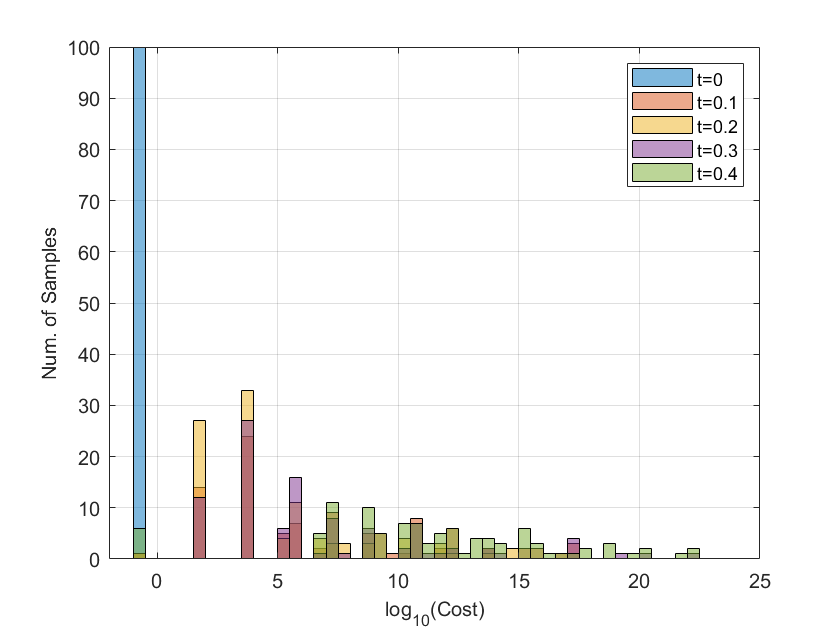}
    \caption{BK distillation cost applied to states after one-axis twisting $|00000\rangle$ initial states with random misalignment. Here $\theta_{\rm max}=0.05$. }
    \label{fig:oneaxis_z}
\end{figure}

First of all, squeezing alone is not the key to improving the distillation outcome --- clearly a state needs to contain enough magic to be distillable at all. Although squeezing in general can add magic to a stabilizer state, a slight squeezing of an initial state without spin misalignment does not improve distillation outcome. Furthermore, states that are unsqueezed under the same unitary $U(t)$ (eqn \ref{eqn:oneaxis}) also sees improvement in distillation. For example, distillation outcome is just as good, if not better, for initial states where $\bar{\theta}_i =0$ (Figure~\ref{fig:oneaxis_z}). However $U(t)$ does not squeeze $|00000\rangle$, which is approximately the initial state we have.

\begin{figure}%
    \centering
    \subfloat[\centering ]{{\includegraphics[width=0.46\textwidth]{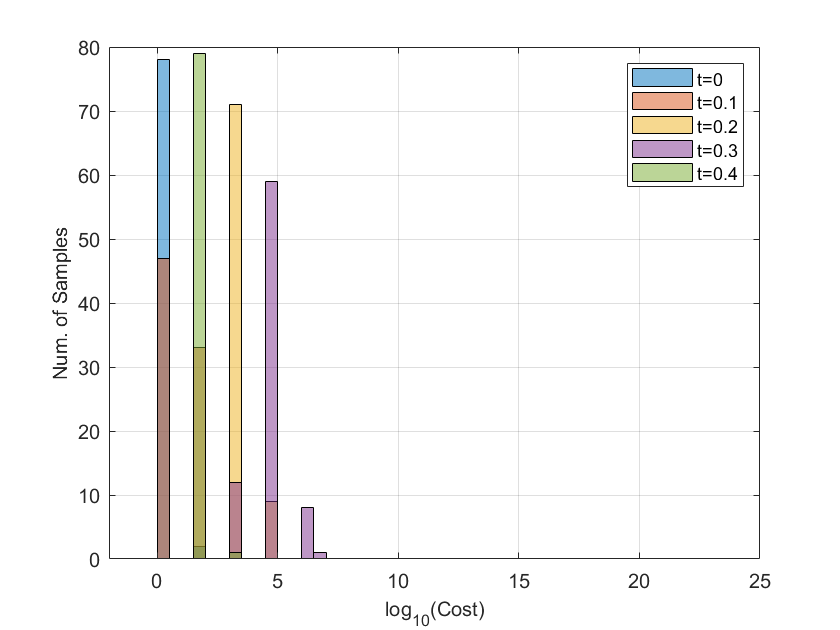} }}%
    \qquad
    \subfloat[\centering ]{{\includegraphics[width=0.46\textwidth]{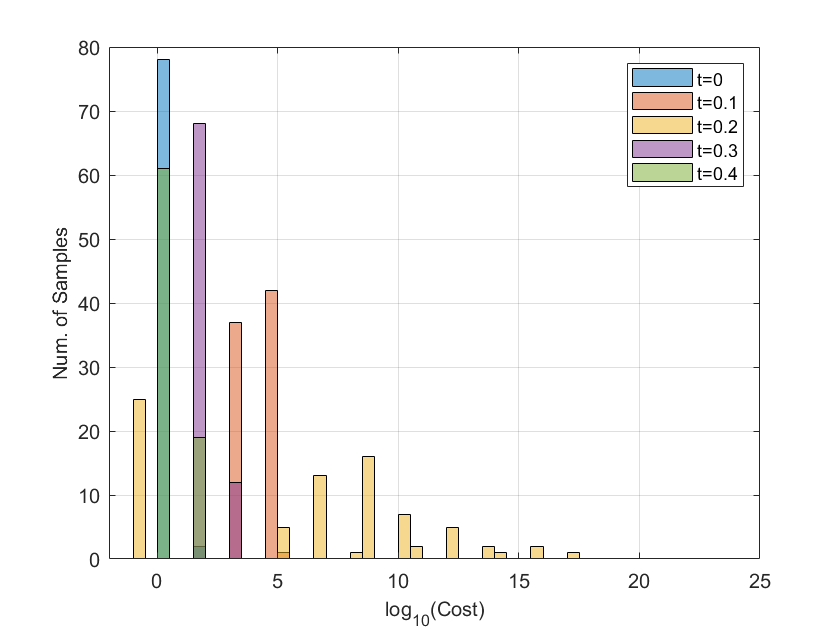} }}%
    \caption{(a) Squeezing the tensor product of five noisy $T$ states via one-axis twisting. (b) Squeezing the tensor product of give noisy $T$ states via two-axis countertwisting. Here $\beta_{\rm max}=0.05, p_{\rm max}=0.1$ for both plots. Squeezing increases the overall cost of distillation in both cases. }%
    \label{fig:nearT_distill}%
\end{figure}

The same global unitary action $U(t)$ also need not improve distillation for other types of initial states. Sometimes it acts as coherent noise to the system, which impedes distillation. For example, consider the scenario where the individual qubit input states are sufficiently close to $|T\rangle$ such that $\overline{\beta_i} = \theta_T= \arccos(1/\sqrt{3})/2$, $\phi=\pi/4$
\begin{align}
    |\tau_0^i\rangle &= \cos(\beta_i)|0\rangle+\exp(i\phi)\sin(\beta_i)|1\rangle\\
    |\tau_1^i\rangle &= -\sin(\beta_i)|0\rangle +\exp(i\phi)\cos(\beta_i)|1\rangle,
\end{align}
and the initial state is given by 
\begin{equation}
    \rho_{\rm ini} = \bigotimes_i (p_i|\tau_0^i\rangle\langle\tau_0^i|+(1-p_i)|\tau_1^i\rangle\langle\tau_1^i|).
    \label{eqn:msin}
\end{equation}
As usual, $\beta_i$ is drawn uniformly random for an interval $[\overline{\beta_i}-\beta_{\rm max},\overline{\beta_i}+\beta_{\rm max}]$. The probability $p_i$ is also drawn uniformly for an interval $[\overline{p_i}-p_{\rm max},\overline{p_i}+p_{\rm max}]$. These are precisely the target magic $T$ states when $p_{\rm max}, \beta_{\rm max}=0$.
Then squeezing now acts as a source of noise that renders distillation less efficient (Figures~\ref{fig:nearT_distill}).  

\begin{figure}
    \centering
    \includegraphics[width=0.7\textwidth]{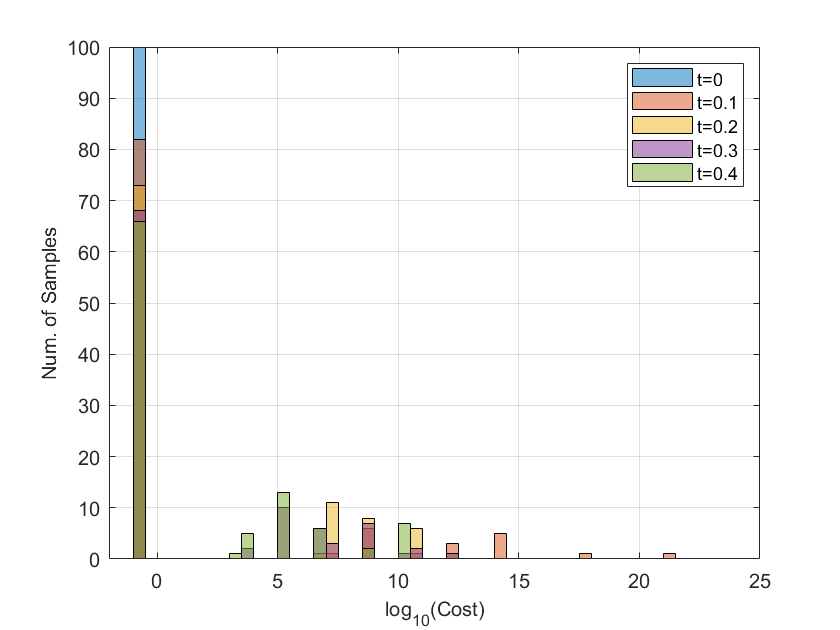}
    \caption{One-axis twisting applied to initial mixed states that are close to a product stabilizer state. $\theta_{\rm max}=0.05$ and $p_{\rm max}=0.02$.}
    \label{fig:mixed_oneaxis}
\end{figure}

On the other hand, if the initial state is close to $|+++++\rangle$ or $|00000\rangle$ but mixed, then the improvement in the distillation cost also quickly diminishes in these examples as a one increases $p_{\rm max}$. For instance, consider states of the form (\ref{eqn:msin}) but with $\phi=0, \bar{\beta}_i=\pi/4$, the distillation cost is shown in Figure~\ref{fig:mixed_oneaxis} where the number of undistillable states can increase sharply even when the purity is still relatively high.

\begin{figure}%
    \centering
    \subfloat[\centering ]{{\includegraphics[width=0.45\textwidth]{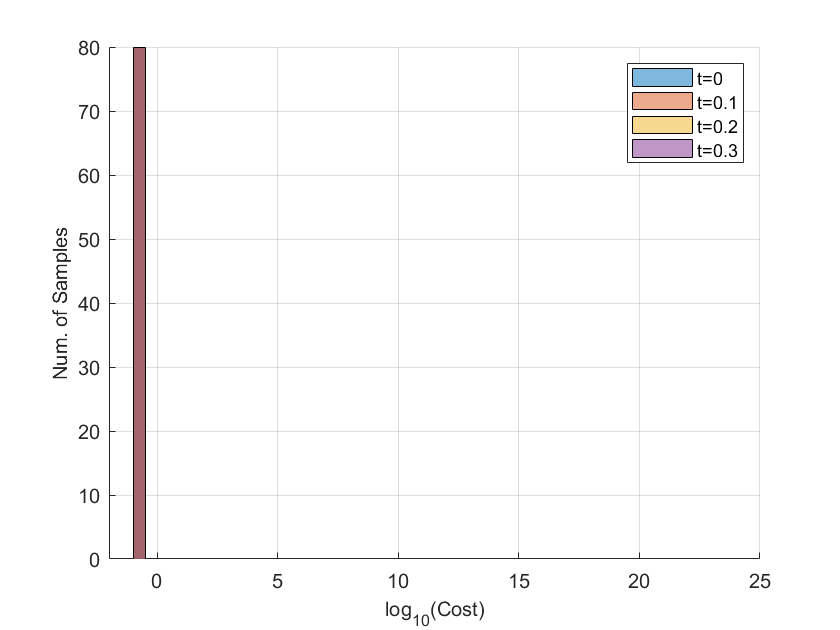} }}%
    \qquad
    \subfloat[\centering ]{{\includegraphics[width=0.45\textwidth]{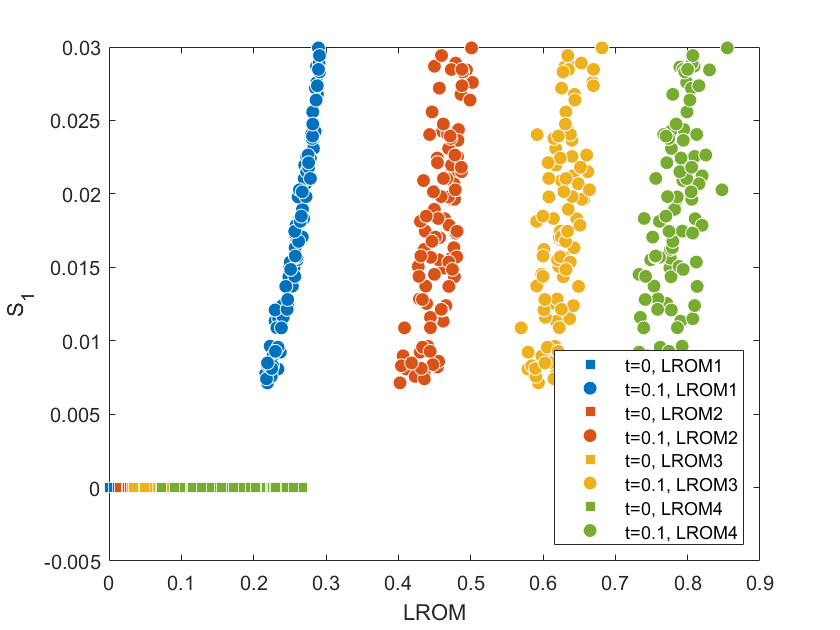} }}%
    \caption{(a) None of the states are distillable after two axis countertwisting. (b) LROM and single site entanglement of the state with (colored disks) and without squeezing (colored squares). }%
    \label{fig:two_axis_plot}%
\end{figure}

Now let us briefly examine the same process but with a different global unitary applied to the $\bar{\theta}_i=\pi/4$ initial state~(\ref{eqn:mis_spin_state}). Consider a different squeezing procedure using two-axis countertwisting~\cite{SSS93} with $V(t)=\exp(-i t H)$ where 
\begin{equation}
H= \frac {1}{2i}(S_+^2-S_-^2),~~ S_{\pm} = S_x \pm iS_y.
\end{equation}
Then squeezing produces a similar level entanglement as the previous one-axis twisting procedure for similar values of $t$.  It also increases the total magic in the state. However, no visible enhancement for the distillation can be found (Figure~\ref{fig:two_axis_plot}). 

\begin{figure}
    \centering
    \includegraphics[width=0.7\textwidth]{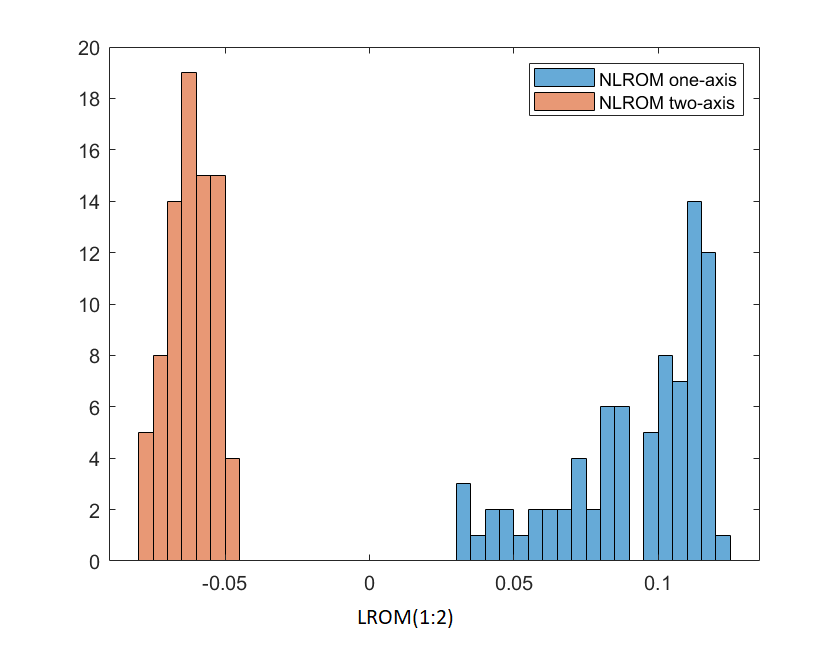}
    \caption{Histogram contrasting two-body nonlocal magic $\mathcal{LROM}(1:2)$ after the one-axis twisting and the two-axis countertwisting options that result in similar levels of entanglement ($S_1 \sim 0.08$ and $S_1\sim 0.02$ respectively). Slight negativity can be attributed to the non-additivity of LROM.}
    \label{fig:NLROM}
\end{figure}

The lack of improvements here may be attributed to the amount of non-local magic that is added to the system. Note that a pure state with non-local magic is most certainly entangled, but an entangled state need not contain non-local magic. While there is a slight addition to non-local magic using the one-axis twisting unitary, there is virtually no change in the two-axis countertwisting scheme (Figure~\ref{fig:NLROM}). We estimate  non-local magic using 
\begin{equation}
    \mathcal{LROM}(1:2)=\log(\mathcal{R}(\rho_{12}))-[\log(\mathcal{R}(\rho_{1}))+\log(\mathcal{R}(\rho_2))],
\end{equation}
where $\rho_{12}$ is an arbitrarily chosen two-site subsystem.


\end{document}